\documentclass[aps,prb,twocolumn,showpacs,preprintnumbers,amsmath,amssymb]{revtex4}
\usepackage{graphicx}% Include figure files \usepackage{bm}% bold

\begin{document}

\title{Low-$T$ Phononic Thermal Conductivity in Superconductors with Line Nodes}
\author{M. F. Smith} \affiliation{National Synchrotron Research Center, Nakhon Ratchasima, 30000 Thailand.}
\date{\today}

\begin{abstract}
The phonon contribution to the thermal conductivity at low temperature in superconductors 
with line nodes is calculated assuming that scattering by both 
nodal quasiparticles and the sample boundaries is significant.  It is determined that, within the regime in 
which the quasiparticles are in the universal limit and the 
phonon attenuation is in the hydrodynamic limit, there exists a wide temperature range over which the phonon thermal conductivity varies as $T^2$.  This behaviour comes from the 
fact that transverse phonons propagating along certain directions do not 
interact with nodal quasiparticles and is thus found to be required by the symmetry of the 
crystal and the superconducting gap, independent of the model used for 
the electron-phonon interaction.  The $T^2$-dependence of the phonon thermal conductivity occurs over a well-defined intermediate temperature range: at higher $T$ the temperature-dependence is found to be linear while at lower $T$ the usual $T^3$ (boundary-limited) behaviour is recovered.   Results are compared to recent measurements of the thermal conductivity of Tl2201, and are shown to be consistent with the data.   
\end{abstract}


\maketitle

Measurements of the thermal conductivity $\kappa$ in cuprates and other unconventional 
superconductors, made in the milliKelvin temperature range, 
 have been valuable in studying T=0 properties of quasiparticles near the nodes of 
the gap.  The measured heat current is carried by both
electrons and phonons, and it is often difficult to identify their respective contributions.  
To isolate the electronic contribution, the low-$T$ data is typically 
fit to a smooth curve that is extrapolated (eg. Ref. \onlinecite{suth03} and references therein) to T=0.  Assuming that the extrapolated value  
is dominated by electrons, the electronic part $\kappa_{el}/T (T=0)$ 
can be obtained without knowing the temperature-dependence of the phononic 
contribution.  This procedure has yielded
values of $\kappa_{el}/T (T=0)$ that are consistent with one another and with independent experimental 
and theoretical results\cite{huss02}\cite{chia00}\cite{meso99}\cite{durs00}\cite{suzu02}\cite{sun95}.  However, if one wishes to obtain any 
temperature dependence of $\kappa_{el}/T$, then some understanding of the 
phononic contribution is needed.  This is particularly true when scattering of phonons by electrons is 
important, for in this case the thermal currents cannot easily be disentangled by analyzing the 
magnetic field dependence of the data.  It is also in this case that the observed phonon 
heat current may be used to study the electrons. 

The importance of electron-phonon interactions in the low-$T$ phononic thermal conductivity has been implicated in recent 
measurements\cite{hawt05} on Tl$_2$Ba$_2$CuO$_{6+\delta}$ (Tl2201).  The impurity bandwidth $\gamma$ is estimated  to be 10-40 K for this 
material\cite{hawt_priv}, so the measured temperature range of below 1 K falls within the 
universal limit and $\kappa_{el}/T$ is expected to be independent of temperature.  The data 
can be fit very well by an expression: 
\begin{equation}
\kappa/T=\alpha+\beta T 
\end{equation}
where $\alpha$ and $\beta$ are independent of $T$.  The phononic thermal conductivity $\kappa_{ph}$ appears to be 
proportional to $T^2$ (instead of $T^3$, as expected for boundary-limited scattering).  Earlier measurements\cite{behn99}\cite{movs98} on overdoped BSCCO and recent studies\cite{hawt_priv} of LSCO show
similar behaviour.  The authors of Ref. \onlinecite{hawt05} 
attributed the $\beta T$ term to phonons scattered by electrons, which is one 
obvious candidate for the relevant mechanism.

Motivated by these data and by the general issues mentioned above, I have calculated the thermal conductivity of phonons in the universal regime under the assumption 
that the significant current relaxation processes are scattering from nodal quasiparticles and scattering from sample 
boundaries.  The main result of this article is the following: due to the presence of inactive phonon modes (i.e. modes for 
which the matrix element with nodal quasiparticles is zero), there occurs a wide temperature 
range within the 
universal regime over which the phononic thermal conductivity 
$\kappa_{ph}$ varies as $T^2$.  The width of this temperature range is such that the 
temperature may vary by a factor as much as $\Delta_0/\gamma$, which 
can be between 10 and 400 for optimally doped cuprates.  The $\kappa_{ph}\propto T^2$ behaviour is expected to occur within a well-defined intermediate temperature range, with $\kappa_{ph}\propto T^3$ predicted at lower temperatures (i.e. in the usual boundary-scattering limit) and $\kappa_{ph}\propto T$ predicted at higher temperatures.  Below, I use experimental values for material-parameters to estimate the location of the temperature crossovers and find that they are consistent with the observation of $T^2$ behaviour from below 100 mK up to several Kelvin in Tl2201.  I have also compared the predicted magnitude of $\beta$ to the value obtained from the Tl2201 data and found good agreement.

Later in this article, I use a particular model for the electron-phonon interaction to calculate $\kappa_{ph}$ when scattering by quasiparticles is important.  
The main qualitative result (the existence of a region in which $\kappa_{ph}$ 
varies as $T^2$) does not depend on the model chosen, but follows from symmetry and 
other general considerations as I will first show.   

Symmetry arguments can be used to prove that the electron-phonon matrix element 
between certain acoustic phonon states and the quasiparticles at the gap nodes is zero  (such phonon states are called `inactive'\cite{more96}). 
For example, it has been shown\cite{walk01} that in tetragonal crystals with a 
$d_{x^2-y^2}$ gap, a transverse phonon having its polarization vector in the basal plane (the CuO$_2$ plane in 
cuprates) is inactive when the basal-plane component of its wavevector is parallel to the (110) direction.  
The interaction between inactive phonons and quasiparticles is obtained by expanding the matrix element, which 
varies with quasiparticle momentum on the scale of the Fermi wavevector $k_f$,
about the nodal points.  In the universal limit, quasiparticles occupy a strip of the Fermi 
surface of length $k_f(\gamma/\Delta_0)$ centred on 
each node.  The electron-phonon coupling for inactive phonons is 
thus smaller than that for active phonons by a factor $\gamma/\Delta_0$ and the corresponding phonon lifetime 
(proportional to the inverse-square of the matrix element) larger by $\Delta_0^2/\gamma^2$.  This can have 
a significant effect on low-$T$ phonon thermal transport, which I now consider.
    
Using a relaxation-time Boltzmann equation description of the phonons, assuming that Mathiesson's law 
holds for the combined effects of scattering from quasiparticles and from the sample boundaries and 
ignoring any anisotropy in the speed of sound $c_s$, we have
\begin{equation}
\kappa_{ph} = C (k_B T)^3 \int dx x^4 \big{(-}\frac{\partial n}{\partial x}\big{)} \int d\Omega \frac{m^2(\theta,\phi)}{1+(\Lambda/c_s)\tau_{ph}^{-1}}
\label{boltz}
\end{equation}
where $\tau^{-1}_{ph}$ is the scattering rate of phonons by quasiparticles, $\Lambda$ is the mean free path for 
boundary scattering, i.e. the length of the sample, $C=\Lambda c_s^{-2} k_B(2\pi\hbar)^{-3}$, $n(x)$ is 
a Bose function and $m(\theta,\phi)$ is the usual
angular factor that gives heaviest weight to phonons moving parallel to the temperature gradient.  This
factor is not important unless the temperature gradient is applied perpendicular to the wavevectors of all inactive phonon states; it may be ignored in the 
present discussion.  The first integral is over a dimensionless variable corresponding to phonon energy (only $x$ between 1 and 10 contribute much to the integral), the 
second is over phonon direction.  

For sufficiently low phonon frequency, $\tau_{ph}^{-1}$ can be described 
in the hydrodynamic approximation, where it varies quadratically with phonon frequency 
and can be written as
\begin{equation}
(\Lambda/c_s)\tau_{ph}^{-1}=x^2(T/T_0)^2f^2(\phi,\theta)
\label{hydro_tau}
\end{equation}
where $T_0$ is a temperature scale and $f(\phi,\theta)$ is proportional to the 
electron-phonon matrix element.  

If the matrix element is roughly isotropic, as expected for longitudinal 
modes, then the result for $\kappa_{ph}$ will clearly be a function that crosses over at $T_0$ from $T^3$ 
at low $T$ (when boundary scattering, dominates) to linear-$T$ at higher $T$ (when quasiparticle 
scattering does).  This does not adequately describe the $T$ dependence of $\kappa_{ph}$ in the latter 
case, since the main contribution comes from transverse modes for which $f^2(\theta,\phi)$ is highly 
anisotropic.    

When scattering by quasiparticles is dominant, the current is carried by the phonons that suffer the least 
attenuation by quasiparticles, i.e. inactive phonons.  One may thus consider only modes for which states having particular directions of propagation 
will be inactive.  Even for these modes, a typical direction of propagation will result in active attenuation 
(in the example given above, a transverse phonon polarized in the basal plane that propagates along [100] 
is active, so the lifetime of a phonon in this mode can be expected to decrease by a 
factor $\gamma^2/\Delta_0^2$ as it is turned away from [110]).  If the planar component (i.e. component perpendicular to the line node) of the wavevector of an inactive phonon makes an angle $\phi_0$ with the $\hat{\bf x}$-axis then $\tau^{-1}_{ph}$ increases rapidly with $\phi-\phi_0$.  The dependence on $\theta$, measured along the node, is comparatively 
weak.    

There is a temperature region of width 
$\Delta_0/\gamma$ over which the lifetime of active (inactive) phonons will be 
limited by scattering from quasiparticles (sample boundaries).  In other words, a region over which the
second term in the denominator of Eq. \ref{boltz} is negligible for $\phi=\phi_0$ and large for typical $\phi$.  Within this 
temperature region, the factor $1/[1 + (\Lambda/c_s)\tau_{el}^{-1}]$ will be given by
\begin{equation}
\frac{1}{1+x^2(T/T_0)^2f^2(\theta,\phi)}\approx\frac{T_0}{xT}\frac{\pi}{|\partial f(\theta,\phi)/\partial \phi|}\delta(\phi-\phi_0)
\end{equation}
so that 
\begin{equation}
\kappa= C (k_B T)^2 (k_B T_0) \Gamma[4]\zeta[3]\pi \int d\cos(\theta) n^2(\theta,\phi_0)
\label{main}
\end{equation}
where I have combined the functions of angle into $n^2(\theta,\phi)$.  The evolution of $\kappa_{ph}$ with temperature is thus:  $T^3$ at the lowest temperature, crossing over to $T^2$ at $T_0$ (evaluated for an active phonon) then finally crossing over to 
linear-in-$T$ at a temperature $T_0 (\Delta_0/\gamma)$.  

It should be noted that this result is valid in the hydrodynamic 
limit of the phonon attenuation, in which the phonon-frequency dependence is quadratic.  The origin of the 
$T^2$-dependence is therefore unrelated to the usual $T^2$ behaviour of this quantity that is seen in 
normal metals, which occurs when the attenuation is in the quantum 
limit and has a corresponding linear-dependence on phonon frequency\cite{pipp}\cite{butl78}.    
The $T^2$ regime is expected in the usual case for clean superconductors with line nodes, that is 
whenever inactive phonon states exist and $\gamma/\Delta_0<<1$.  

Having established the existence of a $T^2$ range in $\kappa_{ph}$, I will now consider a particular model for the electron-phonon 
interaction in order to determine whether this $T^2$ behaviour is likely to be observed under realistic conditions.  This is done for 
the case relevant to cuprates so that a comparison may be made with the Tl2201 data.               

A form for the electron-phonon interaction in tight-binding square lattice materials such as cuprates 
is described in Ref. \onlinecite{walk01}.  The dependence of the matrix element on quasiparticle and 
phonon wavevector (${\bf k}$ and ${\bf q}$, respectively) are obtained for acoustic modes with the coupling strength characterized by the energy $g=a|dt/dR|$, the derivative of the 
hopping parameter $t$ with respect to bond length in the effective single-band Hamiltonian for the 
CuO$_2$ plane ($a$ is the lattice constant).  Using this and, based on the discussion above, 
considering only transverse phonons polarized in the plane, the matrix element is  
\begin{equation}
|M({\bf k},{\bf \hat{q}})|=\frac{(g\omega_{\bf q})}{MNc_s^2} g({\bf k},\hat{\bf q})
\label{mel}
\end{equation}
where
\begin{equation}
g({\bf k},\hat{\bf q})=\sin(\theta)\bigg{[}\frac{\Delta_{\bf k}}{\Delta_0} \sin(2\phi)+\frac{g^{\prime}}{g}\cos(2\phi)\bigg{]},    
\label{gform}
\end{equation}
$M$ is the mass of the unit cell and $N$ is the number of copper sites in the crystal.  The angles are defined such that $(\theta,\phi)=(\pi/2,\pi/4)$ describes a phonon travelling along (110), so $\phi_0=\pi/4$.  Also, $g^{\prime}\equiv 4\sqrt{2}a|\partial t^{\prime}/\partial R|\sin^2(k_fa)$, 
where $k_f$ is the Fermi wavevector at the node.  The gap has been introduced 
using $\Delta_{\bf k}/\Delta_0= \cos(k_x a) - \cos(k_y a)$ to describe the ${\bf k}$-dependence 
of the nearest-neighbour matrix element.  Note that Eq. \ref{mel} has the symmetry-imposed
property mentioned in the previous section: it is zero for $\phi=\pi/4$ and $k_x=k_y$.  The first (second) term in Eq. 6 describes inactive (active) phonon states.  

The inverse-lifetime of phonons is equal to twice the magnitude of the imaginary part of the retarded
 phonon self-energy evaluated at the real phonon frequency $\omega_{\bf q}=c_s |{\bf q}|$.  I use the lowest approximation 
for the phonon self-energy, given by the `bare-bubble' diagram with the electron-Green's functions 
taken in the universal limit\cite{durs00}\cite{samo} and the vertex given by Eq. \ref{mel} so that
\begin{equation}
\tau_{ph}^{-1}= \frac{8}{\pi} \frac{g^2}{NMc_s^2}\sum_{\bf k} \frac{\omega_{\bf q}^2\gamma^2 g^2({\bf k},{\bf \hat{q}})}{(E_{\bf k+q}^2+\gamma^2)(E_{\bf k}^2+\gamma^2)}  
\label{lifetime}
\end{equation}
where I have used $k_B T, \omega_{\bf q} << \gamma$ and $\hbar=1$.

For active phonon states, the matrix element is independent of ${\bf k}$ and can be moved outside the summation.  
The contribution to the sum over ${\bf k}$ that comes from the vicinity of each node is given by
\begin{equation}
\sum_{\bf k} \frac{\omega_{\bf q}^2\gamma^2}{(E_{\bf k+q}^2+\gamma^2)(E_{\bf k}^2+\gamma^2)}=\frac{n_0}{2 v_2 k_f}\bigg{[}\frac{\omega_{\bf q}^2}{\tilde{\omega}_{\bf q}^2}\gamma^2 S_A(\tilde{\omega_{\bf q}}/2\gamma)\bigg{]} 
\label{actsum}
\end{equation}
where
\begin{equation}
S_A(x)=\frac{2}{R(x)}\ln\bigg{(}\bigg{|}\frac{1+R(x)}{1-R(x)}\bigg{|}\bigg{)},
\end{equation}
$R(x)=\sqrt{1-x^{-2}}$, $n_0$ is the normal density of states, $v_2 k_f \approx \Delta_0$, and $\tilde{\omega}_{\bf q}$ is
 the magnitude of the wavevector $\tilde{\bf q}\equiv v_f {\bf q}\cdot{\bf {\hat k_1}}+v_2 {\bf q}\cdot{\bf {\hat k_2}}$, 
with ${\bf \hat{k_1}}$, and ${\bf \hat{k_2}}$ unit vectors directed perpendicular, parallel to the 
Fermi surface at the node, respectively.   

In the hydrodynamic limit\cite{walk01} $(\tilde{\omega}_{\bf q}/2\gamma\to 0)$, the factor in square brackets in Eq. \ref{actsum} is equal to $\omega_{\bf q}^2$.  
For large $\tilde{\omega}_{\bf q}/\gamma$ (a limit valid\cite{smit03} only when $c_sq<<\gamma<<v_2q$) this factor becomes 
$4\frac{\gamma^2}{v_2^2}\ln(\omega_{\bf q}/\gamma)$.  

For the inactive phonon state obtained by taking $\theta=\pi/2,\phi=\pi/4$ above
so that $g^2({\bf k},{\bf \hat{q}})=\Delta^2_{\bf k}/\Delta_0^2$, the corresponding result is
\begin{equation}
\sum_{\bf k} \frac{(\Delta^2_{\bf k}/\Delta_0^2)\omega_{\bf q}^2\gamma^2}{(E_{\bf k+q}^2+\gamma^2)(E_{\bf k}^2+\gamma^2)}=\frac{n_0}{2v_2k_f}\bigg{[}\frac{\omega_{\bf q}^2}{\tilde{\omega}_{\bf q}^2}\gamma^2 S_I(\tilde{\omega_{\bf q}}/2\gamma)\bigg{]}
\label{insum}
\end{equation}
with
\begin{equation}
S_I({\bf x})=\frac{2x^2\gamma^2}{\Delta_0^2}\bigg{[}\ln\bigg{(}\frac{\Delta_0^2}{\gamma^2}\bigg{)}+\big{[}1-R+R\bigg{(}\frac{x_2}{x}\bigg{)}^2\big{]}S_A(x)\bigg{]}
\end{equation}
where $\tilde{q_2}$ is the ${\bf k}_2$ component.  A phonon propagating with $\phi=\pi/4$ has $\tilde{q}_2=0$
for two of the nodes and $|\tilde{q_2}|=\tilde{q}$ for the other two.  The latter give the main contribution 
to the sum over nodes since $S_I$ and $S_A$ are monotonically decreasing with 
$E_{{\bf k}+{\bf q}}$ and $v_f >> v_2$ in most cases.  
In the hydrodynamic limit the factor in square brackets in Eq. \ref{insum} is $(\gamma^2/\Delta_0^2)\ln(\Delta_0/\gamma)\omega_{\bf q}^2$.  This is the result noted above that, for phonons propagating in an inactive direction ($\phi=\pi/4$), the magnitude of $\tau^{-1}_{ph}$ is smaller than for typical directions by a factor $\gamma^2/\Delta_0^2$.  

To obtain $\kappa_{ph}$ at temperatures well below $T(\Delta_0/\gamma)$, I use the hydrodynamic limit for active phonons in Eq. \ref{boltz}, ignore $(\Lambda/c_s)\tau_{ph}^{-1}$ for inactive phonons
compared to unity
and take the temperature gradient along the plane with $m^2(\theta,\phi)=\sin^2(\theta)\cos^2(\phi)$
and $f^2(\theta,\phi)=\cos^2(2\phi)\sin^2(\theta)$. The result is 
\begin{equation}
\kappa_{ph}=C(k_BT)^3\int dx x^3 \bigg{(}-\frac{\partial n}{\partial x}\bigg{)} s(x\tau)
\label{tanmess}
\end{equation}
where $\tau=T/T_0$, $s(x)=\frac{\pi}{x^2}[1-(x^{-1}-x)\mathrm{atan}(x)]$, and
\begin{equation}
(k_BT_0)^{-2}=\frac{2}{\pi}\frac{\Lambda}{c_s v_2 k_f}\bigg{(}\frac{g^{\prime 2} n_0}{N M c_s^2}\bigg{)}.  
\label{tnot}
\end{equation}
Eq. \ref{tanmess} describes the smooth crossover, occuring at $T\approx T_0$, from $T^3$ to $T^2$ behaviour in $\kappa_{ph}$.  In the limit $T_0<<T<<T_0(\Delta_0/\gamma)$, Eq. \ref{tanmess} gives Eq. \ref{main}, with the integral in the latter equal to $\pi/2$.  Note that both $T_0$ and $\beta$, the magnitude of the $T^2$ term in $\kappa_{ph}$, are universal, i.e. independent of $\gamma$.  Also, they do not depend on the detailed angle-dependence of the electron-phonon matrix element, but only on the fact that $\tau_{ph}^{-1}$ has a deep minimum at $\phi=\phi_0$ (see Ref. \onlinecite{Tlinearnote}).  

For a numerical estimate, I use $g\approx 6$ eV obtained from observations of low-$T$ downturns\cite{smit05} of $\kappa_{el}/T$ in optimally doped cuprates and assume that $g^{\prime}$ is of the same order\cite{srru} as $g$; the factor in parentheses of 
Eq. \ref{tnot} is then of order 
unity and, introducing the Debye energy $\epsilon_D$, $k_BT_0\approx\sqrt{\Delta_0 \epsilon_D (a/\Lambda)}$.  
For $\Lambda$ equal to 0.3 mm (a value typical of the cross-sectional width of samples 
used in the measurements\cite{hawt05}\cite{prou02}), $T_0$ is roughly 0.3 K and the coefficient $\beta$ is 0.5 mW K$^{-3}$ cm$^{-1}$.  
The former value indicates that it is plausible that the $T^2$ region may be observed within the sub-Kelvin $T$-range, 
and the latter value is within a factor of two of that measured for Tl2201.  Also, using\cite{hawt_priv} $\gamma=3$ meV and $\Delta_0$ = 40 meV, 
the linear-in-$T$ behaviour would be expected to occur above 4 K (see Ref. \onlinecite{new}).      

The model gives similar predictions for other cuprate materials (doping dependence 
is considered below), whereas many data on cuprates\cite{suth03}\cite{chia00}\cite{tail96}\cite{take02} indicate that $\kappa_{ph}$ has a 
$T$-dependence that is closer to $T^3$.  However, the estimates used for the relevant parameters 
(especially $g^{\prime}$) are very rough, and the value of $T_0$ obtained
here is such that, within a $T$-range between 50 mK and 1 K or so, neither $T^2$ nor $T^3$ behaviour 
can be ruled out.  This work does provide a basis for understanding $T^2$-dependence in cuprates 
when it occurs and should be considered when, for example, the magnetic-field dependence of 
$\kappa$ is analyzed.  A detailed comparison of this work with the low-$T$ data is complicated by the observation of $T^{3-\delta}$ (where $\delta\le0.3$) power laws arising because of a poorly understood boundary-scattering effect\cite{suth03}\cite{tail96} and because low-$T$ downturns in $\kappa/T$ data occur due to heating losses\cite{smit05}.  

I shall briefly discuss the doping dependence of these results.  From Ref. \onlinecite{hawt05}, the 
$T^2$ coefficient $\beta$ increases slightly with doping across the overdoped side of the phase 
diagram.  Since $T_0\propto \sqrt{\Delta_0/g^{\prime 2}}$, the 
results above are consistent with this dependence only if $g^{\prime 2}$ decreases with 
overdoping, i.e. if it is roughly proportional to $\Delta_0$.  It is expected
that $g^{\prime}$, being derived from the effective hopping Hamiltonian, should vary with doping and it would 
be interesting to see whether such a doping-dependence can be observed elsewhere.  

$T^2$ behaviour is less likely to be observed for underdoped 
materials.  This is because $v_2/c_s$ becomes very large ($10^2$ or so\cite{suth03}) at extreme underdoping, so the hydrodynamic approximation breaks down at temperatures much less than those for which the universal 
limit is still valid.  (For overdoped samples $v_2/c_s<6$ so the requirement for the hydrodynamic limit, $(v_2/c_s)k_BT<<\gamma$, 
is not much more restrictive than that for the universal limit, $k_BT << \gamma$.)  If $(c_s/v_2)\gamma\le T_0$ then $(c_s/\Lambda)\tau^{-1}_{ph}$ 
would not exceed one until temperatures much higher than $T_0$ and, once it did, $\kappa_{ph}$ would vary as $T^3/\ln(T)$ rather than $T^2$.  This may explain why the $T^2$ dependence in $\kappa_{ph}$ has only been seen for overdoped cuprates.

In summary, I have found that the phononic thermal conductivity $\kappa_{ph}$ varies as $T^2$ over a wide, experimentally accessible temperature range within 
the universal limit.  This work may be useful in 
analyzing low-$T$ thermal conductivity data, especially in studies aimed at determining temperature-dependent 
corrections to the universal value of $\kappa_{el}/T$.         
I thank D. G. Hawthorn, Mike Sutherland and M. B. Walker for useful discussions.


\begin{thebibliography}{99}
\bibitem{Tlinearnote}  In contrast, the $T$-linear term in $\kappa_{ph}$ found for $T>>T_0(\Delta_0/\gamma)$ is not universal and depends strongly on the angular-dependence of the matrix element. The linear-term, and the crossover behaviour for $T\approx T_0(\Delta_0/\gamma)$, can be calculated by inserting Eq. \ref{lifetime}, with both active and inactive states included, into Eq. \ref{boltz}.  This is omitted for brevity.
\bibitem{srru} For Sr$_2$RuO$_4$, the value of $g^{\prime}/g$ was found\cite{walk01}\cite{lupi01} to be very small, such that the attenuation of certain active modes
is less than that for inactive modes in this material, i.e. $g^{\prime}/g < \gamma/\Delta_0$.  I will assume that the Sr$_2$RuO$_4$ case is unusual and that $g^{\prime}/g\approx t^{\prime}/t$ is more typical.
\bibitem{new} New measurements on Tl2201 by D. G. Hawthorn show a crossover from $T^2$ to $T$ behaviour occuring just above 1K, which is in reasonable agreement with my estimate of 4 K given the uncertainty of the latter.  I was informed of these results after submitting this article.
\end{thebibliography}
\end{document}